%% file: Main.tex
\documentclass[conference]{IEEEtran}
\IEEEoverridecommandlockouts
\usepackage{cite}
\usepackage{amsmath,amssymb}
\usepackage{algorithmic}
\usepackage{graphicx}
\usepackage{xcolor}
\usepackage{balance}
\usepackage{acronym}
\usepackage{comment}
\usepackage{multirow}
\usepackage{subcaption}
\usepackage[font=footnotesize]{caption}
\usepackage{eso-pic}

\newcommand{\IEEEcopyrightnotice}{%
  \AddToShipoutPictureBG*{%
    \AtPageLowerLeft{%
      \put(\LenToUnit{0.75in},\LenToUnit{0.25in}){%
        \parbox[b]{\dimexpr\paperwidth-1.5in\relax}{%
          \fontsize{7}{8}\selectfont
          \textcopyright\ 2026 IEEE. Personal use of this material
          is permitted. Permission from IEEE must be obtained
          for all other uses, in any current or future media,
          including reprinting/republishing this material for
          advertising or promotional purposes, or reuse of any
          copyrighted component of this work in other works.
        }%
      }%
    }%
  }%
}

\input{acronyms}

\def\BibTeX{{\rm B\kern-.05em{\sc i\kern-.025em b}\kern-.08em
    T\kern-.1667em\lower.7ex\hbox{E}\kern-.125emX}}
\begin{document}

\bstctlcite{IEEEexample:BSTcontrol}
\title{UE-Side Location Privacy for 5G NR Uplink Positioning: Mechanisms and Trade-offs 
\thanks{This work was supported, in part, by  the  European Research Council (ERC) under the
European Union’s Horizon Europe (Grant agreement No. 101078411); the Swedish Research Council (project 2023-03821), and the Horizon Europe R\&I Programme under Horizon-JU-SNS project PAISES-6G (Grant Agreement No. 101292896).}
}

\author{\IEEEauthorblockN{
Giulia Focarelli\IEEEauthorrefmark{1},
Alireza Pourafzal\IEEEauthorrefmark{2},
Henk Wymeersch\IEEEauthorrefmark{2}, and
Stefania Bartoletti\IEEEauthorrefmark{1}}
\IEEEauthorblockA{\IEEEauthorrefmark{1}Department of Electronic Engineering, University of Rome Tor Vergata, Rome, Italy,\\
}
\IEEEauthorblockA{\IEEEauthorrefmark{2}Department of Electrical Engineering, Chalmers University of Technology, Gothenburg, Sweden,\\
Email: \{giulia.focarelli, stefania.bartoletti\}@uniroma2.it; \{alireza.pourafzal, henkw\}@chalmers.se.}
}

\maketitle
\IEEEcopyrightnotice
\begin{abstract}
Future 5G-Advanced and 6G networks increasingly reuse uplink communication waveforms for positioning and sensing. This raises privacy concerns, as \acp{UE} may unintentionally reveal precise timing information even when positioning services are not explicitly requested. While prior works show generic \ac{OFDM} pilots can be manipulated to degrade \ac{ToA} estimation without compromising data links, this paper extends these concepts to a realistic 5G \ac{NR} uplink framework  including \ac{SRS}, \ac{DMRS}, \ac{PUSCH}, and standardized 3GPP channel models.  We investigate several \ac{UE}-side privacy mechanisms: optimized pilot distortion, artificial noise, artificial multipath, and delay spoofing. Through 3GPP-compliant  sample-level simulations, we assess their impact via localization, communication, and consistency-based detection metrics. The resulting analysis highlights the trade-offs among privacy, communication reliability, and detectability, providing key insights into waveform-level obfuscation for future \ac{ISAC} systems.
\end{abstract}


\section{Introduction}
Mobile networks are no longer used only for connectivity. In 5G \ac{NR}, standardized \acp{RS} such as downlink \ac{PRS}, \ac{SRS}, and \ac{DMRS} enable accurate timing measurements, channel estimation, and positioning services \cite{3gpp.38.211,3gpp.38.305}. This evolution is central to \ac{ISAC}, but it also creates a privacy problem: a \ac{gNB} that receives uplink \acp{RS} may infer the range or location of a \ac{UE} even when the UE does not intend to reveal it. This risk is particularly relevant because the same uplink resources that support reliable communication can also expose high-resolution timing information.
\begin{figure}[!t]
\centering

\includegraphics[width=\linewidth]{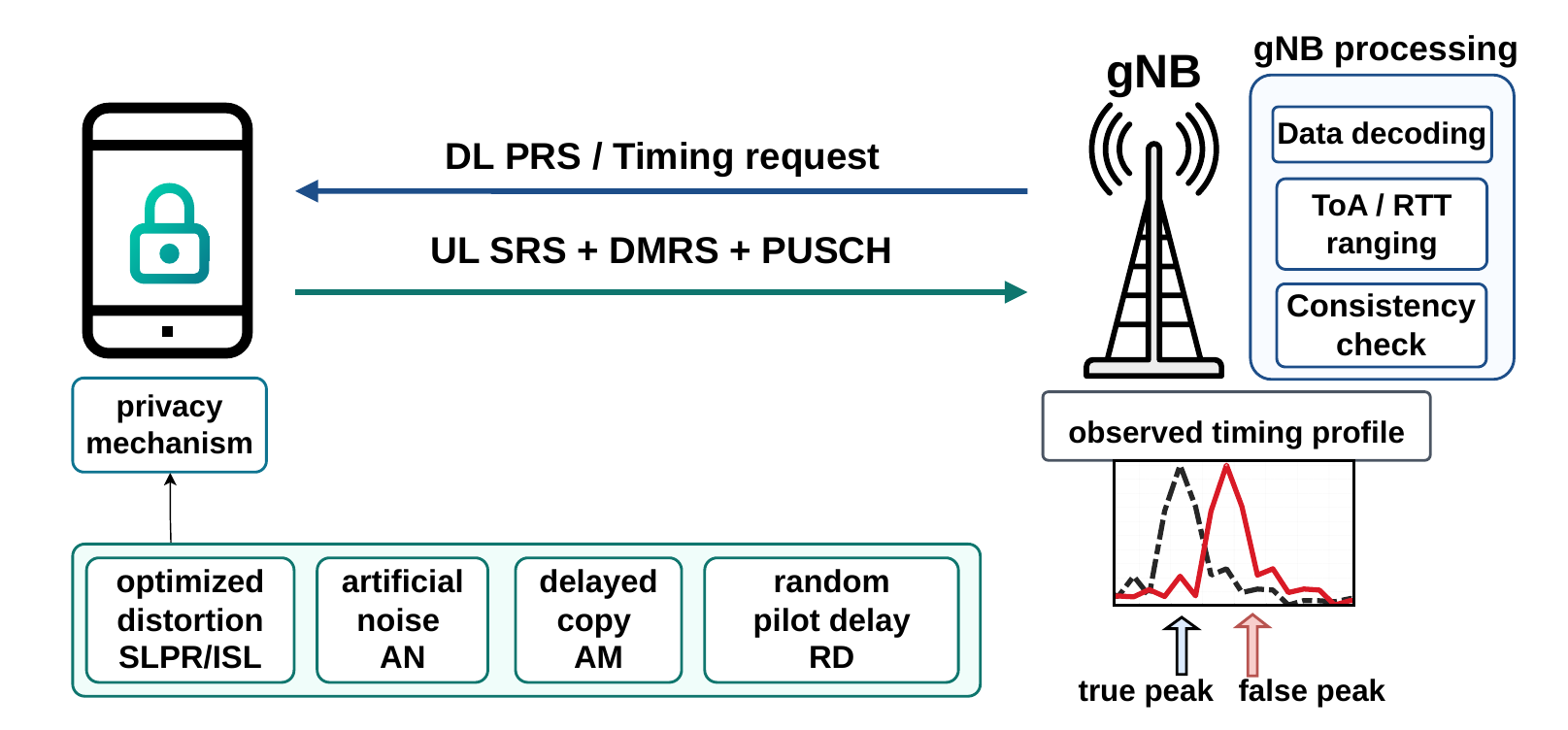}
\vspace{-2mm} 
\caption{Considered scenario: the UE obfuscates its ranging metrics while preserving the data link. The gNB jointly processes the same SRS, DMRS, and PUSCH resources for decoding and ToA/RTT estimation.}
\label{fig:scenario}
\vspace{-1mm}
\end{figure}

Location-privacy mechanisms at the physical layer can broadly be grouped into obfuscation and spoofing methods. Obfuscation methods degrade the quality of localization observables. For instance, \ac{AN} and spatial perturbation have been employed to mask user distance from WiFi access points \cite{ayyalasomayajula2023users} and cellular networks \cite{checa2020location}. More recently, transmit beamforming has been optimized to conceal sensitive location data in MIMO-OFDM \cite{zhang2025privacy} and cross-layer architectures \cite{tomasin2022beamforming}. Spoofing methods, instead, manipulate the physical channel signatures to create a controlled false observation. This can be achieved by injecting \ac{AM} components \cite{li2024channel}, spoofing delay and angle features \cite{li2025delay}, or dynamically altering the spatial signatures in mmWave environments \cite{italiano2025holotrace}. Recent attacks on 5G positioning, including targeted manipulations such as selective spoofing \cite{gao2024surgical} as well as blind full-frame replay attacks \cite{BarFocPalZanMelBleBia,FocZanPalBiaBar:26}, have also shown that unauthenticated or manipulated reference-signal measurements can mislead positioning and sensing procedures. While these works strongly motivate \ac{UE}-side signal-level defenses, they do not directly answer how much privacy can be gained in a realistic 5G \ac{NR} uplink while strictly preserving the communication link.
The closest basis for this work is the generic \ac{OFDM} pilot-distortion framework in \cite{ercan2025pilot}, where a UE applies a complex per-subcarrier distortion vector to increase sidelobes of a mismatched ambiguity function. That paper introduced two useful design criteria, the \ac{SLPR} and the \ac{ISL}, and showed that delay estimation can be degraded with limited communication loss in an idealized OFDM setting. 

This work investigates how effectively \ac{UE}-side waveform manipulations can obfuscate unintended uplink ranging in 5G \ac{NR} while preserving reliable communication. Specifically, our contributions are threefold: (i) we formulate the unintended ranging problem where the \ac{gNB} exploits both \ac{SRS} and \ac{DMRS} timing observables; (ii) we adapt optimized pilot-distortion techniques to \ac{NR} standards and compare them with \ac{AN}, \ac{AM}, and delay spoofing; and (iii) through a 3GPP-compliant link-level framework, we systematically characterize the fundamental trade-offs among location privacy, communication penalty, and resistance to network-side consistency checks.

\section{System Model}
We consider an uplink single-anchor scenario consisting of a single-antenna \ac{UE} and a single-antenna \ac{gNB} operating in a 5G \ac{NR} \ac{OFDM} network. The \ac{UE} is located at an unknown three-dimensional position $\mathbf{p}_{\text{UE}} \in \mathbb{R}^3$, while the \ac{gNB} is deployed at a known coordinate $\mathbf{p}_{\text{gNB}} \in \mathbb{R}^3$. The physical geometric distance between the transmitter and the receiver is defined as $d = \|\mathbf{p}_{\text{UE}} - \mathbf{p}_{\text{gNB}}\|$. The \ac{gNB} jointly performs data demodulation and \ac{ToA} estimation on the received signals. Due to clock offsets, this one-way \ac{ToA} serves as the uplink component of a \ac{RTT} measurement; to isolate the effect of \ac{UE}-side privacy mechanisms, the downlink timing is assumed error-free.

\subsection{5G Uplink Reference Signals}
According to the \ac{3GPP} 5G \ac{NR} physical-layer specifications \cite{3gpp.38.211}, the uplink time-frequency resource grid contains both information-bearing physical channels and \acp{RS}. Specifically, for the uplink transmission we consider the following components: 
\begin{itemize}
    \item \text{\ac{PUSCH}:} Carries the
user data. Its transmission format is determined by the \ac{MCS}, which specifies the modulation order and coding rate.
    \item \text{\ac{DMRS}:} A reference signal associated with the
    \ac{PUSCH} allocation, used by the \ac{gNB} for channel estimation
    and coherent equalization of the received data symbols.
    \item \text{\ac{SRS}:} Uplink positioning reference signal featuring wideband configurability and good correlation properties, enabling precise \ac{ToA} measurements at the \ac{gNB}.
\end{itemize}

\subsubsection{Demodulation Reference Signals}
The uplink \ac{DMRS} is a \ac{PUSCH}-associated \ac{RS} generated as a pseudo-random QPSK sequence according to
\cite{3gpp.38.211}, with initialization depending on the \ac{DMRS}
scrambling identity, the slot number, and the configured antenna port. Each element of the generated sequence is mapped to physical \ac{RE} $(k,l)$
inside the scheduled \ac{PUSCH} allocation, where $k$ denotes the
subcarrier index and $l$ the \ac{OFDM} symbol index within the slot.
The time-domain locations are determined by the \ac{PUSCH} mapping type
(Type A or Type B), the allocation duration $L_{\mathrm{DMRS}}$, and the configured additional
\ac{DMRS} positions. Under mapping Type A, the first \ac{DMRS} symbol index $l_0$ is determined by higher-layer parameters. Under mapping Type B, the \ac{DMRS} symbol indices are referenced to the beginning of the \ac{PUSCH} allocation, with the first \ac{DMRS} symbol located at $l_0=0$.

\subsubsection{Sounding Reference Signals}
The \ac{3GPP} defines the \ac{SRS} as the uplink \ac{RS} for timing-based positioning procedures. In the uplink localization procedure, the \ac{UE} transmits the 
\ac{SRS}, while one or multiple \acp{gNB} process the received waveform to
estimate timing observables. Unlike \ac{DMRS}, which is associated with data demodulation, positioning
\ac{SRS} is configured to sound the uplink channel over a potentially wide
frequency span, enabling high-resolution delay-profile estimation. The
\ac{SRS} sequence is obtained from cyclic shifts of a base sequence according to \cite{3gpp.38.211}.
The sequence is mapped to \ac{RE} $(k,l)$ within the \ac{OFDM}
resource grid. The mapping depends on the transmission comb
$K_{\mathrm{TC}}\in\{2,4,8\}$, which determines the frequency-domain comb spacing, the bandwidth configuration parameters
$C_{\mathrm{SRS}}$ and $B_{\mathrm{SRS}}$, and the number of consecutive
\ac{SRS} symbols $L_{\mathrm{SRS}}\in\{1,2,4\}$. These parameters are
configured through higher-layer signaling according to the UE and network
capabilities.

\subsection{Signal and Channel Model}
Let $N_{\text{RB}}$ denote the number of Resource Blocks (RBs) within the carrier bandwidth, where each RB comprises $N_{\text{sc}}^{\text{RB}} = 12$ subcarriers. The total number of subcarriers is $K = N_{\text{RB}} \times N_{\text{sc}}^{\text{RB}}$, with a subcarrier spacing (SCS) $\Delta f = 2^{\mu} \times 15\text{ kHz}$. 
The frequency-domain transmitted resource grid over a given slot is represented by a matrix $\mathbf{X} \in \mathbb{C}^{K \times L}$, where $L$ is the number of \ac{OFDM} symbols per slot. At subcarrier index $k \in \{0, \dots, K-1\}$ and symbol index $l \in \{0, \dots, L-1\}$, the grid element $X[k,l]$ is populated according to the standard resource element mapping rule:
\begin{equation}
X[k,l] =
\begin{cases}
x_{\mathrm{data}}[k,l],
& (k,l) \in \mathcal{I}_{\mathrm{data}}, \\

a_{\mathrm{DMRS}}[k,l],
& (k,l) \in \mathcal{I}_{\mathrm{DMRS}}, \\

a_{\mathrm{SRS}}[k,l],
& (k,l) \in \mathcal{I}_{\mathrm{SRS}},
\end{cases}
\label{eq:defX}
\end{equation}
where $\mathcal{I}_{\mathrm{data}}$, $\mathcal{I}_{\mathrm{DMRS}}$,
and $\mathcal{I}_{\mathrm{SRS}}$ denote mutually disjoint sets of
time-frequency resource elements carrying PUSCH data symbols, \ac{DMRS},
and \ac{SRS}, respectively. For the $l$-th \ac{OFDM} symbol, and omitting the fixed symbol index for
notational simplicity, the transmitted baseband signal is modeled as
\begin{equation}
    s(t)=
    \sum_{k=0}^{K-1}
    X[k] e^{j2\pi f_k t},
    \quad 0\leq t<T.
    \label{eq:ofdm_mod_ul}
\end{equation}
where $f_k$ denotes the baseband frequency of the $k$-th subcarrier,
with spacing $\Delta f$, and $T=1/\Delta f$ is the useful \ac{OFDM}
symbol duration.
The complete transmitted waveform $s(t)$ is obtained by concatenating the cyclic-prefixed OFDM symbols over all considered slots.
Considering a multipath propagation scenario, the received baseband
signal can be modeled as
\begin{equation}
    r(t) =
    \sum_{n=1}^{N_{\mathrm{p}}}
    \alpha_n s(t-\tau_n)
    + w(t),
    \label{eq:multipath}
\end{equation}
where $N_{\mathrm{p}}$ is the number of propagation paths, while
$\tau_n$ and $\alpha_n$ denote the delay and complex gain of the
$n$-th path, respectively. The term $w(t)$ represents additive receiver
noise.
\subsection{Receiver Processing and Timing Estimation}

At the \ac{gNB}, the received uplink waveform is processed using locally generated replicas of the configured \acp{RS}. Nominally, within the 5G \ac{NR} framework, the \ac{SRS} and the \ac{DMRS} serve different primary objectives: the \ac{SRS} is generally allocated for uplink waveform synchronization and \ac{ToA} estimation, whereas the \ac{DMRS} is intended for channel estimation and coherent data demodulation. However, in a location-privacy violation context, the \ac{gNB} can opportunistically exploit the standardized and fully known \ac{DMRS} sequence as an alternative positioning pilot, especially when the \ac{SRS} is obfuscated. To address this, the receiver processing is formulated for a generic pilot type $p \in \{\mathrm{SRS},\mathrm{DMRS}\}$. Let $\tilde{a}_{p}(t)$ denote the time-domain replica associated with a given pilot type $p$, generated at the \ac{gNB} from
the nominal reference symbols and their configured resource-element
mapping. The receiver computes the cross-correlation between the
received waveform and the local pilot replica over a finite observation
window.
Let $r[n]=r(nT_s)$ and $\tilde{a}_{p}[n]=\tilde{a}_{p}(nT_s)$ denote the
sampled received signal and the sampled local pilot replica,
respectively, with
sampling time $Ts$. The discrete correlation profile is
$
    R_p[q] =
    \sum_{m \in \mathcal{W}_p}
    r[m] \tilde{a}_{p}^{*}[m-q],
$
where $\mathcal{W}_p$ denotes the observation window used for the
correlation. Under multipath model in \eqref{eq:multipath}, this
correlation is
$
    R_p[q]
    =
    \sum_{n=1}^{N_{\mathrm{p}}}
    \alpha_n \mathring{R}_{p}[q-q_n]
    + \eta_p[q],
$
where $q_n=\lfloor \tau_n/T_s \rfloor$ is the sampled delay of the
$n$-th path, $\eta_p[q]$ is the filtered noise term, and
$\mathring{R}_{p}[q]$ is the autocorrelation of the nominal pilot
replica when the transmitted and locally generated pilots coincide. In this work, the timing estimate associated with pilot type $p$ is
obtained using the maximum-correlation rule
$\hat{q}_p = \arg\max_q |R_p[q]|$,
and the corresponding delay estimate is
$\hat{\tau}_p = \hat{q}_p T_s$.

Within the standardized framework for single-anchor RTT positioning, the uplink procedure relies exclusively on the \ac{SRS}. The \ac{gNB} transmits a downlink \ac{PRS} to the \ac{UE}; after reception, the \ac{UE} responds in the uplink with the \ac{SRS} after a processing time $T_{\mathrm{proc}}$. The measured
round-trip time is
$
    T_{\mathrm{RTT}}
    =
    \hat{\tau}_{\mathrm{DL},\mathrm{prs}}
    +
    T_{\mathrm{proc}}
    +
    \hat{\tau}_{\mathrm{UL},\mathrm{srs}},
$
where $\hat{\tau}_{\mathrm{DL},\mathrm{prs}}$ is the downlink timing estimate,
$\hat{\tau}_{\mathrm{UL},\mathrm{srs}}$ is the uplink timing estimate obtained from
\ac{SRS}, and $T_{\mathrm{proc}}$ is the reply time due to processing computed at
the \ac{UE} device and it is assumed to be either included in the uplink transmission payload or known a priori by the \ac{gNB}.
After subtracting
the nominal processing time, the RTT-based range estimate is
$
    \hat{d}
    =
    \frac{c}{2}
    \left(
    T_{\mathrm{RTT}}
    -
    T_{\mathrm{proc}}
    \right).
$
Although the \ac{DMRS} is not configured as a positioning signal and does not define a \ac{RTT} procedure by itself, its deterministic and
standardized structure can still be exploited by a gNB-side observer. In
particular, it can serve as an auxiliary timing observable, acting as a
consistency check against \ac{SRS} manipulation and, in multi-anchor
settings, as a possible fallback when the \ac{SRS} is obfuscated.
\subsection{Privacy Model}
We assume an honest-but-curious \ac{gNB}, trusted for data demodulation but untrusted for localization. The \ac{UE} is assumed to know the nominal uplink transmission it is scheduled to send, while the \ac{gNB} processes the received signal using nominal reference-signal replicas. Privacy mechanisms are implemented directly at the \ac{UE} by intentionally modifying the transmitted reference resources. All mechanisms satisfy the same average transmit-power constraint, ensuring that privacy gains do not arise from increased uplink power.

\section{Privacy Mechanisms}
\label{sec:privacy_mechanisms}

This section details the specific waveform-level manipulations applied by the \ac{UE} to obfuscate ranging observables. To prevent the network from easily bypassing the manipulation, the considered techniques are applied to both \ac{SRS} and \ac{DMRS}. We investigate four distinct approaches, ranging from structured phase distortions to unstructured noise injection.
\begin{figure*}[t]
    \centering

    \begin{subfigure}[t]{0.195\textwidth}
        \centering
        \includegraphics[width=\linewidth]{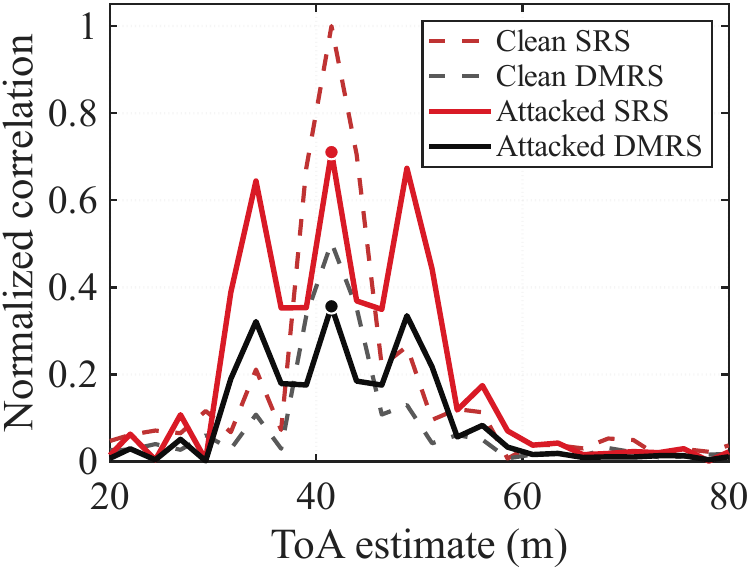}
        \caption{SLPR,  $\epsilon = 0.5$}
        \label{fig:corr_slpr}
    \end{subfigure}
    \hfill
    \begin{subfigure}[t]{0.195\textwidth}
        \centering
        \includegraphics[width=\linewidth]{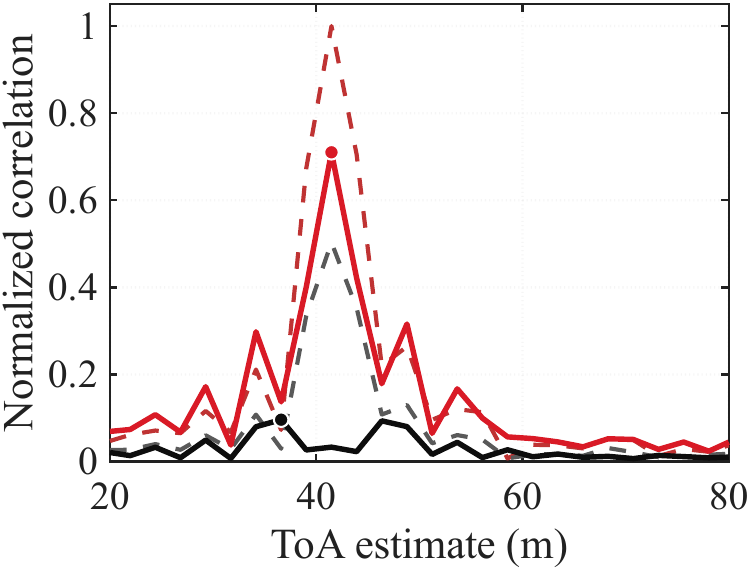}
        \caption{ISL,  $\epsilon = 0.5$}
        \label{fig:corr_isl}
    \end{subfigure}
    \hfill
    \begin{subfigure}[t]{0.195\textwidth}
        \centering
        \includegraphics[width=\linewidth]{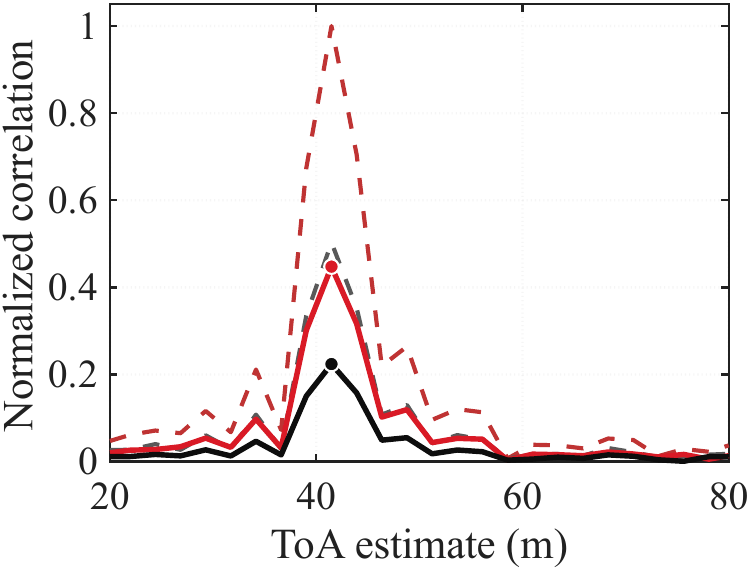}
        \caption{AN, $\sigma_{\mathrm{AN}}=10$}
        \label{fig:corr_an}
    \end{subfigure}
    \hfill
    \begin{subfigure}[t]{0.195\textwidth}
        \centering
        \includegraphics[width=\linewidth]{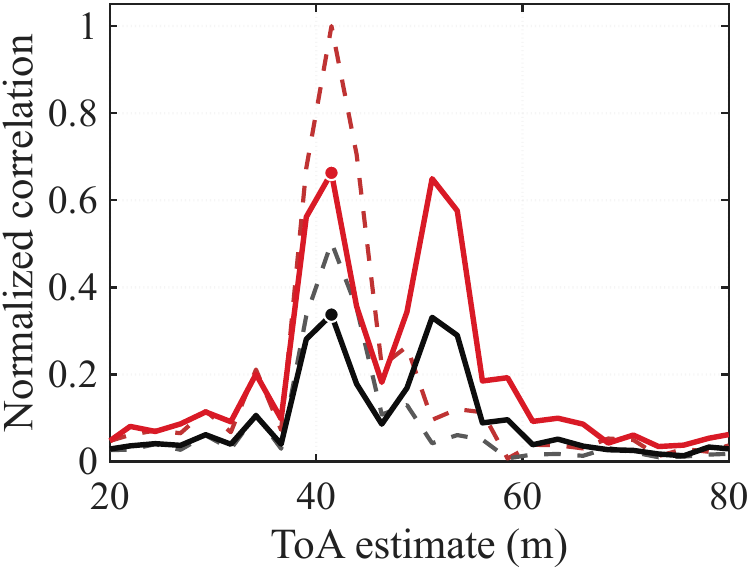}
        \caption{AM,  $|\gamma_1|^2 = 0.5$}
        \label{fig:corr_am}
    \end{subfigure}
    \hfill
    \begin{subfigure}[t]{0.195\textwidth}
        \centering
        \includegraphics[width=\linewidth]{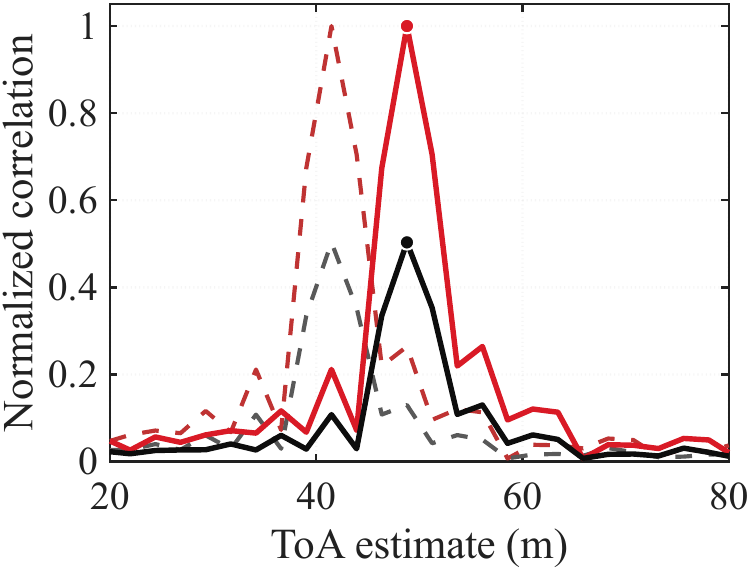}
        \caption{RD, $d_{\max} = 10$\,m}
        \label{fig:corr_random_delay}
    \end{subfigure}

    \caption{Representative LOS timing-correlation profiles at fixed UE position. For each mechanism, the clean SRS and DMRS correlations are shown with dashed lines, while the obfuscated SRS and DMRS correlations are shown with solid lines.}
    \label{fig:correlation_profiles}
\end{figure*}
\subsection{Optimized Pilot Distortion}
The first mechanism is the optimized pilot distortion of
\cite{ercan2025pilot}, adapted to the 5G NR \ac{SRS} and \ac{DMRS}. Let $\mathcal K_p$ denote the set of occupied pilot
subcarriers for pilot type $p$, with
relative frequencies $f_k$. The \ac{UE} applies a complex
frequency-domain coefficient $z_p[k]$ to each pilot-bearing subcarrier,
so that each pilot on subcarrier $k\in\mathcal K_p$ carries
$z_p[k]a_p[k]$ instead of the nominal $a_p[k]$. The \ac{gNB} correlates the received signal with the nominal pilot replica, creating a mismatched profile that may shift the dominant peak from the true delay.

The delay-dependent mismatched correlation term can be written as
$
    \chi_p(\Delta\tau;\mathbf z_p)
    =
    \sum_{k\in\mathcal K_p}
    |a_p[k]|^2 z_p[k] e^{j2\pi f_k \Delta\tau}.
$
where $\Delta\tau$ is the delay offset relative to the true timing peak. The clean pilot corresponds to $\mathbf z_p=\mathbf 1$, for which the largest correlation peak is expected at $\Delta\tau=0$. Pilot distortion selects $\mathbf z_p$ to increase sidelobe energy while keeping the transmitted pilot close to the communication-friendly clean signal. A compact form of the design problem is $\max_{\mathbf z_p} f_{\mathrm{loc}}(\mathbf z_p)$ subject to the constraints $\|\mathbf z_p-\mathbf 1\|^2 \leq \epsilon |\mathcal K_p|$ and $\|\mathbf z_p\|^2 \leq |\mathcal K_p|$, where $\epsilon$ controls the distortion severity, and the objective function $f_{\mathrm{loc}}$ is selected according to the desired ranging-obfuscation criterion. We consider two ranging-obfuscation  criteria. The \ac{SLPR}-based design increases the dominant sidelobe relative to the mainlobe, making a wrong delay peak more likely. The \ac{ISL}-based design increases the total sidelobe energy, spreading ambiguity over a wider delay region. In both cases, the distortion severity is controlled by a parameter $\epsilon$, which limits the distance between the distorted pilot and the communication-optimal pilot. In the two cases, $f_{\mathrm{loc}}$ is instantiated either as
\begin{align}
    f_{\mathrm{SLPR}}(\mathbf z_p)
    &=
    \frac{\max_{\Delta\tau\in\mathcal T_{\mathrm{SL}}}
    |\chi_p(\Delta\tau;\mathbf z_p)|^2}
    {|\chi_p(0;\mathbf z_p)|^2},\\
    f_{\mathrm{ISL}}(\mathbf z_p)
    &=
    \frac{\int_{\mathcal T_{\mathrm{SL}}}
    |\chi_p(\Delta\tau;\mathbf z_p)|^2\,\mathrm d\Delta\tau}
    {|\chi_p(0;\mathbf z_p)|^2},
    \label{eq:slpr_isl_metrics}
\end{align}
with $\mathcal T_{\mathrm{SL}}$ denoting the sidelobe delay region outside the nominal mainlobe.

\subsection{Artificial Noise Injection}
The second mechanism adds a random perturbation to the pilot-bearing \ac{RE} and then renormalizes the modified pilots to maintain the baseline total energy, following the \ac{AN} approach for delay-based privacy in \cite{zhang2024privacy}. Let $\mathbf a_p$ denote the vector collecting the pilot symbols
$a_p[k,l]$ over all resource elements $(k,l)\in\mathcal I_p$. For pilot vector $\mathbf{a}_p$, the UE transmits
\begin{equation}
    \mathbf{a}_{p,\mathrm{AN}}
    =
    \sqrt{E_p}
    \frac{\mathbf{a}_p+\sigma_{\mathrm{AN}}\mathbf{u}}
    {\|\mathbf{a}_p+\sigma_{\mathrm{AN}}\mathbf{u}\|},
    \label{eq:artificial_noise_pilot}
\end{equation}
where $\mathbf{u}$ has independent complex Gaussian entries, $E_p=\|\mathbf{a}_p\|^2$, and $\sigma_{\mathrm{AN}}$ controls obfuscation severity. This baseline is simple and standard-compliant at the power level, but its random nature may average out across multiple frames and may visibly degrade channel-estimation consistency.

\subsection{Artificial Multipath or Delayed-Copy Injection}
The third mechanism creates an artificial delayed copy of the uplink pilot, which is the NR-link-level counterpart of \ac{AM} or fake-path studied in \cite{zhang2024privacy,li2024channel}. The \ac{UE} applies a delay $\Delta_{\mathrm{AM}}$ in
the frequency domain through a phase rotation on each pilot-bearing subcarrier.
The transmitted pilot for a fixed OFDM symbol can be written as
\begin{equation}
    a_{p,\mathrm{AM}}[k]
    =
    \left( \gamma_0 + \gamma_1 e^{-j2\pi f_k\Delta_{\mathrm{AM}}} \right) a_p[k],
    \label{eq:artificial_multipath}
\end{equation}
where $\gamma_0$ and $\gamma_1$ control the relative weights of the
nominal and delayed pilot components, with normalization $|\gamma_0|^2+|\gamma_1|^2=1$ when the two components are approximately orthogonal. The severity parameters are the artificial-path delay $\Delta_{\mathrm{AM}}$ and relative power $|\gamma_1|^2$. This method is expected to create a more interpretable false peak than random noise.

\subsection{Random Pilot-Delay Spoofing}

The \ac{UE} applies a random linear phase rotation across the 
pilot-bearing subcarriers, corresponding to an apparent delay of the pilot
waveform. The transmitted pilot on subcarrier k is
\begin{equation}
    a_{p,\mathrm{RD}}[k]
    =
    a_p[k] e^{-j2\pi f_k \Delta_{\mathrm{RD}}},
    \label{eq:random_delay_freq}
\end{equation}
where the delay is randomly drawn as
$
    \Delta_{\mathrm{RD}}
    \sim
    \mathcal{U}
    \left(
    \frac{d_{\min}}{c},
    \frac{d_{\max}}{c}
    \right).
$
The severity is controlled by the maximum delay \(d_{\max}\). Because $\Delta_{\mathrm{RD}}$ varies independently across transmissions, the induced offset does not appear as a fixed deterministic bias, making it more difficult for a single \ac{gNB} to remove it through simple temporal averaging or fixed-offset calibration.

From the viewpoint of an RTT-based ranging procedure, this physical random pilot-delay (RD) shift is equivalent to introducing an apparent reply-time
offset in the uplink response, without explicitly modeling higher-layer timing reports.

\section{Performance Metrics}
\label{sec:metrics}

The performance of the proposed privacy mechanisms is evaluated through metrics describing localization and privacy, consistency-check metrics, and communication performance. Taken together, these metrics provide a basis for characterizing the resulting privacy-communication trade-offs.

\subsection{Localization and Privacy Metrics}
For trial $n$, pilot type $p$, and mechanism $m$, let $\hat{d}_{p,m}^{(n)}$ be the inferred distance and let $d^{(n)}$ be the geometric UE--gNB distance. The signed distance error is
$e_{p,m}^{(n)}=\hat{d}_{p,m}^{(n)}-d^{(n)}$.
We report the empirical bias $b_{p,m} = \frac{1}{N}\sum_{n=1}^{N} e_{p,m}^{(n)}$, the standard deviation $\sigma_{p,m} = [\frac{1}{N-1}\sum_{n=1}^{N}(e_{p,m}^{(n)}-b_{p,m})^2]^{1/2}$, and the $\mathrm{RMSE}_{p,m} = [\frac{1}{N}\sum_{n=1}^{N}(e_{p,m}^{(n)})^2]^{1/2}$.
Bias and standard deviation capture distinct privacy behaviors: delay spoofing mainly induces systematic errors, whereas pilot noise primarily increases estimation variability. Peak-selection behavior is measured directly from the correlation profile. In this case, a clean reference case is used as an implementation-dependent baseline to quantify the impact of the privacy mechanisms. Let $\hat{q}_{p,m}^{(n)}$ be the selected peak index and $\hat{q}_{p,0}^{(n)}$ the clean-baseline peak index for the same UE position and channel realization. The probability of preserving the clean peak and the probability of inducing a peak switch are $P_{\mathrm{corr},p,m} = \Pr\!\left(\hat{q}_{p,m}=\hat{q}_{p,0}\right)$, and $P_{\mathrm{sw},p,m} = 1-P_{\mathrm{corr},p,m}$.

\begin{figure*}[t]
    \centering

    \begin{subfigure}[t]{0.245\textwidth}
        \centering
        \includegraphics[width=\linewidth]{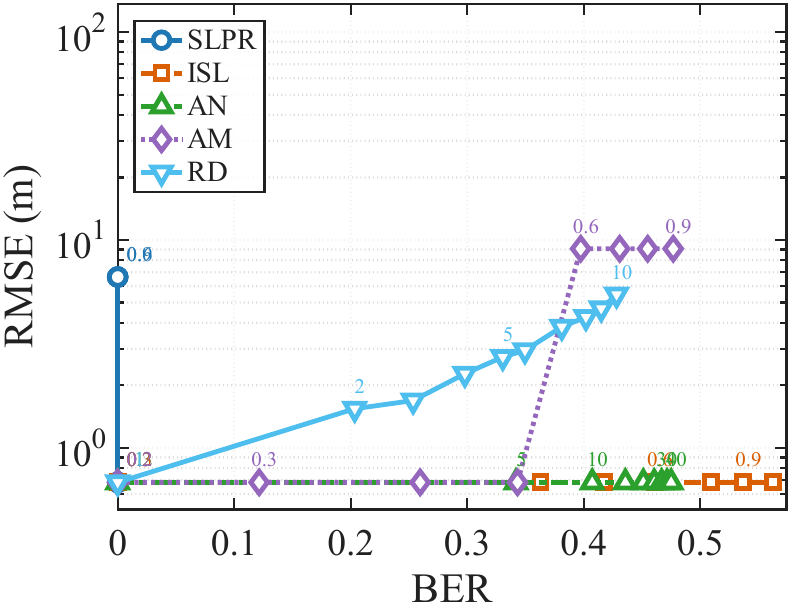}
        \caption{LOS, SRS}
        \label{fig:corr_los_srs}
    \end{subfigure}
    \begin{subfigure}[t]{0.245\textwidth}
        \centering
        \includegraphics[width=\linewidth]{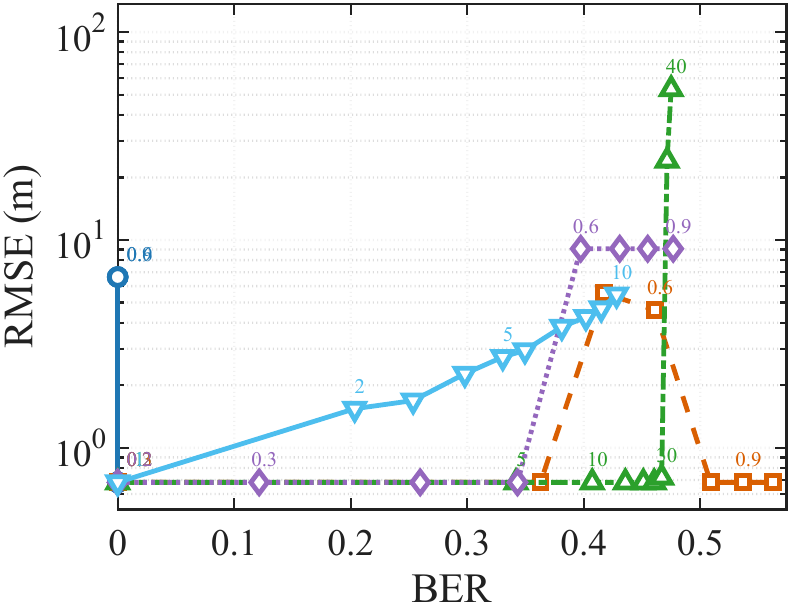}
        \caption{LOS, DMRS}
        \label{fig:corr_los_dmrs}
    \end{subfigure}
    \begin{subfigure}[t]{0.245\textwidth}
        \centering
        \includegraphics[width=\linewidth]{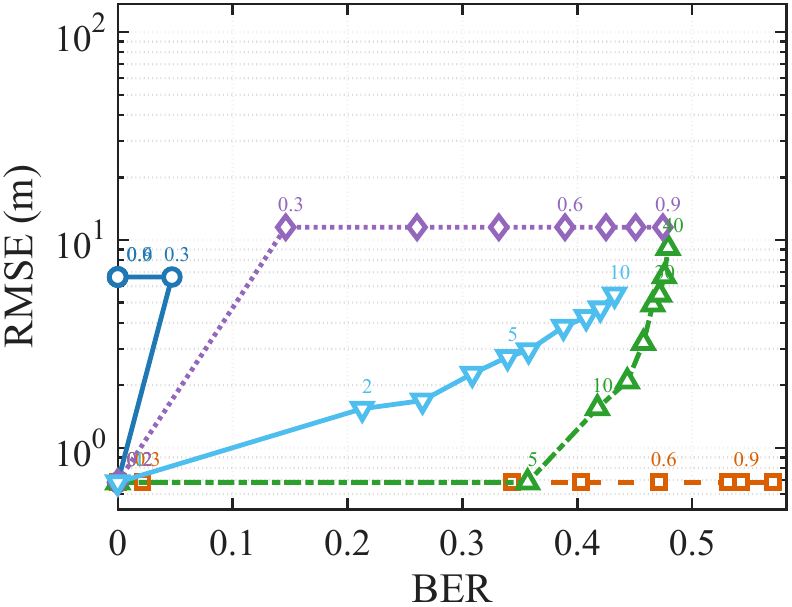}
        \caption{NLOS, SRS}
        \label{fig:corr_nlos_srs}
    \end{subfigure}
    \begin{subfigure}[t]{0.245\textwidth}
        \centering
        \includegraphics[width=\linewidth]{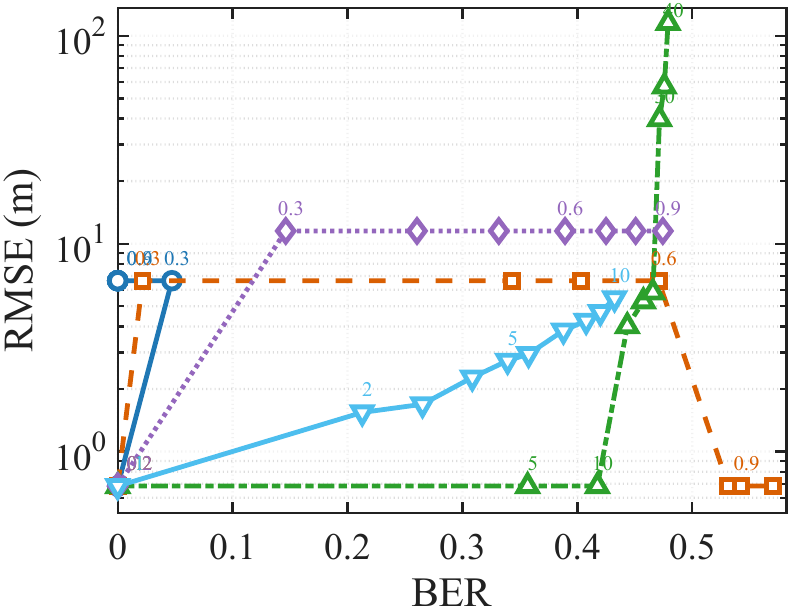}
        \caption{NLOS, DMRS}
        \label{fig:corr_nlos_dmrs}
    \end{subfigure}

    \caption{Privacy-communication trade-off evaluated as positioning \ac{RMSE} versus data \ac{BER}. The performance of the considered \ac{UE}-side manipulation mechanisms is compared for both \ac{SRS} and \ac{DMRS} under \ac{LOS} and \ac{NLOS} propagation conditions. Markers along the curves indicate increasing values of the respective configuration parameters (e.g., manipulation severity).}
    \label{fig:privacy-com-tradeoff}
\end{figure*}
\subsection{Consistency-Check Metrics}
We investigate lightweight \ac{gNB}-side consistency checks to assess the detectability of the proposed privacy mechanisms under network integrity monitoring. For each realization, a test statistic $T_m$ is evaluated against an empirical threshold $\eta(P_{\mathrm{FA}})$ calibrated over the N clean baseline realizations for a target false-alarm probability $P_{\mathrm{FA}}$, yielding the detection probability $
P_{\mathrm{D},m}(P_{\mathrm{FA}})
=
\Pr\!\left(T_m>\eta(P_{\mathrm{FA}})\mid m\right)
$. Specifically, $T_m$ measures the timing discrepancy between the \ac{SRS} and \ac{DMRS} estimates. Since both signals share the same physical channel under legitimate conditions, this check effectively exposes privacy mechanisms that distort the two waveforms differently.

\subsection{Communication Metrics}
\label{subsec:communication_metrics}
Communication performance is evaluated after normal uplink receiver processing. The primary metrics are \ac{BER} and \ac{BLER}. For a given frame, the error rates are computed as
 $\mathrm{BER} = N_{\mathrm{err}}/N_{\mathrm{bits}}$ and $\mathrm{BLER} = N_{\mathrm{blk,err}}/N_{\mathrm{blk}}$,
where $N_{\mathrm{err}}$ and $N_{\mathrm{bits}}$ denote the number of erroneous and total transmitted bits, respectively, while $N_{\mathrm{blk,err}}$ and $N_{\mathrm{blk}}$ represent the number of erroneous and total transport blocks.

\section{Link-Level Performance Evaluation}
\label{sec:simulation_results}
This section defines the link-level 5G NR evaluation used to compare the privacy mechanisms. All metrics are evaluated over a fixed \ac{UE}--\ac{gNB} range and identical channel realizations for both the clean and manipulated cases, enabling manipulation-induced changes to be precisely quantified through paired comparisons.

\subsection{Simulation Setup}
The simulator uses a MATLAB 5G NR link-level setup with PUSCH data, DMRS, and positioning-oriented SRS. The baseline configuration is $\Delta f=30~\mathrm{kHz}$, $N_{\mathrm{RB}}=162$, approximately $60~\mathrm{MHz}$ occupied bandwidth, one transmit antenna, one receive antenna, 16-QAM PUSCH with LDPC coding, DMRS mapping type A, and $L_{\mathrm{DMRS}}=1$. The SRS uses one antenna port, transmission comb $K_{\mathrm{TC}}=4$, $C_{\mathrm{SRS}}=42$, $B_{\mathrm{SRS}}=0$, no frequency hopping, and $L_{\mathrm{SRS}}=4$.
Both \ac{SRS} and \ac{DMRS} are configured to span the full allocated bandwidth. This represents a worst-case scenario for the manipulating \ac{UE}: a wideband \ac{DMRS} provides the \ac{gNB} with a high-resolution timing observable that could easily serve as a reliable fallback. Consequently, to prevent the network from simply bypassing the \ac{SRS} manipulation, all considered mechanisms are applied simultaneously to both pilots. Since \ac{PUSCH} equalization relies on the manipulated \ac{DMRS}, any \ac{DMRS} obfuscation directly degrades the data link and causes a communication penalty. Two propagation conditions are considered: \ac{LOS} TDL-E and \ac{NLOS} TDL-B, both with $30~\mathrm{ns}$ delay spread, combined with the 3GPP Urban Macro path-loss model. The UE is placed at $42$\,m from the gNB, and the UE
transmit power is set to $10~\mathrm{dBm}$. To strictly isolate the impact of the \ac{UE}-side manipulation, the simulated distance metrics evaluate the one-way uplink ranging error, explicitly omitting the \ac{RTT} halving factor. For the \ac{AM}, the relative delay $\Delta_{\mathrm{AM}}$ is selected to yield an equivalent distance $d_{\mathrm{AM}} = c\Delta_{\mathrm{AM}}=10$\,m.
The obfuscation severity is swept through the mechanism-specific control parameter: \(\epsilon\) for the optimized SLPR/ISL pilot distortions, \(\sigma_{\mathrm{AN}}\) for \ac{AN}, \(|\gamma_1|^2\) for \ac{AM}, and \(d_{\max}\) for RD spoofing.

\subsection{Performance Evaluation}
The \ac{UE}-side privacy mechanisms are analyzed from complementary perspectives. First, we examine their impact on the correlation profiles to gain insight into the underlying waveform-level effects. We then investigate the privacy-communication trade-offs and assess their robustness against a network-side consistency check. Finally, we provide an overall comparison in terms of obfuscation performance, communication cost, detectability, and implementation complexity.

\subsubsection{Waveform-Level Effects}

Fig.~\ref{fig:correlation_profiles} reports representative normalized correlation profiles for a \ac{UE} located at a true distance of $42\,$m under \ac{LOS} conditions. In the absence of manipulations, the \ac{SRS} exhibits a significantly stronger correlation peak than the \ac{DMRS}, owing to its longer temporal duration and the resulting higher processing gain. The manipulated profiles, generated with $\epsilon = |\gamma_1|^2 = 0.5$, $\sigma_{\mathrm{AN}}=10$, and $d_{\max}=10\,$m, reveal markedly different waveform-level effects.
The optimized distortions SLPR and ISL fundamentally reshape the correlation structure. SLPR heavily suppresses the true \ac{ToA} peak and creates prominent symmetric false peaks. By contrast, ISL attenuates the main peak while spreading energy across neighboring delays, thereby raising the sidelobe floor without producing isolated deceptive peaks.
The AN mechanism mainly degrades the correlation quality, drastically reducing the peak amplitude at the true distance for both \ac{SRS} and \ac{DMRS}, without generating coherent false peaks. In contrast, the AM mechanism preserves a weakened peak at the true delay while introducing a distinct secondary peak at the artificial distance (around $52\,$m for $d_{\mathrm{AM}}=10\,$m), effectively emulating a strong multipath component. Finally, the RD mechanism acts as a rigid translation of the entire correlation profile. The genuine peak at $42\,$m disappears and an identically shaped peak emerges at the spoofed delay, demonstrating a coherent time-shift manipulation at the waveform level.

\subsubsection{Privacy--Communication Trade-off}

Fig.~\ref{fig:privacy-com-tradeoff} illustrates the trade-off between location obfuscation, quantified by the positioning \ac{RMSE}, and communication reliability, measured by the \ac{BER}. The results are shown for both \ac{SRS} and \ac{DMRS} under \ac{LOS} and \ac{NLOS} conditions, with markers indicating increasing manipulation severity.
Among the considered mechanisms, the optimized SLPR approach provides the most favorable trade-off, achieving nearly $10\,$m of obfuscation while maintaining a negligible \ac{BER}. This demonstrates its ability to bias the timing estimate without significantly affecting the data symbols.
The AM mechanism can reach similar privacy levels, although its communication cost strongly depends on the propagation environment. Under \ac{LOS} conditions, obtaining a $10$\,m bias requires a strong secondary path and leads to a \ac{BER} around $0.35$. In \ac{NLOS}, the interaction with natural multipath reduces the communication penalty to approximately $0.1$, allowing the same privacy level to be achieved more efficiently. The RD exhibits a proportional trade-off, with larger delays providing increased privacy at the expense of higher \ac{BER}, making large distance manipulations increasingly costly for the communication link.
By contrast, AN and ISL are highly inefficient. Meaningful obfuscation is achieved only through severe signal degradation, driving the \ac{BER} above $0.4$ and effectively causing a communication outage. Moreover, the \ac{DMRS} proves more vulnerable than the \ac{SRS} to such brute-force perturbations because its single-symbol structure provides a lower processing gain than the multi-symbol \ac{SRS} configuration.

\begin{figure}[!t]
    \centering
    \includegraphics[width=\linewidth]{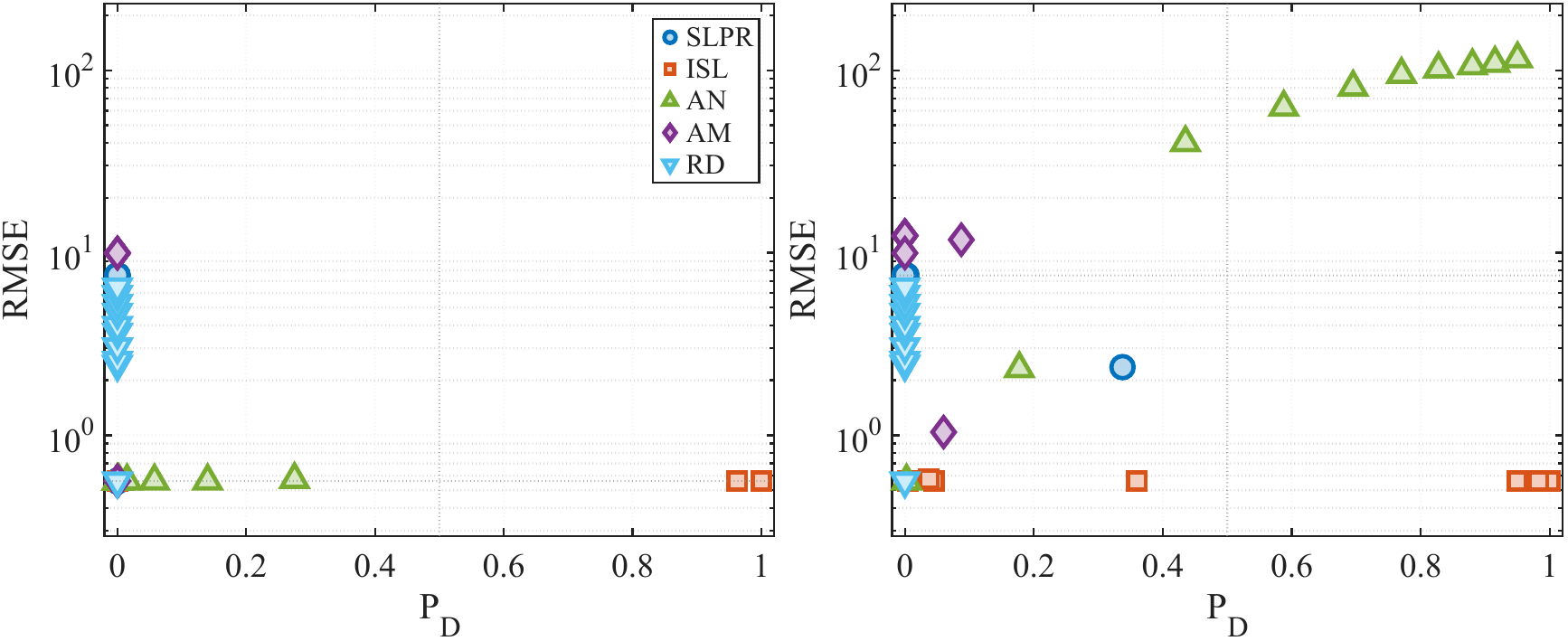}
    \caption{Privacy--consistency trade-off for the considered  mechanisms in LOS conditions (left) and NLOS (right). The detection probability \(P_\text{D}\) is computed from the SRS-DMRS timing disagreement with false-alarm probability \(P_\text{FA}=5\%\).} 
    \label{fig:privacy_consistency_tradeoff}
\end{figure}
\subsubsection{Privacy-Consistency Trade-off}

Fig.~\ref{fig:privacy_consistency_tradeoff} evaluates robustness against a network-side \ac{SRS}-\ac{DMRS} timing consistency check. The detection probability $P_\mathrm{D}$ is computed at a fixed false-alarm probability of $P_\mathrm{FA}=5\%$. An ideal privacy mechanism should combine large positioning errors with low detectability.  A clear distinction emerges between coherent manipulations and unstructured perturbations.  The structured mechanisms  (\ac{SLPR}, AM, RD) shift both \acp{RS} consistently, maintaining a negligible timing disagreement. Thus, they remain highly evasive ($P_\mathrm{D}\approx 0$) even for positioning errors approaching $10\,$m. Conversely, unstructured perturbations (\ac{ISL}, AN) affect the signals independently. The resulting uncorrelated timing errors easily trigger the detector, pushing $P_\mathrm{D}$ towards unity. These results highlight a fundamental limitation: while the consistency check catches incoherent noise, it remains completely blind to coherent waveform manipulations.
\subsubsection{Overall Comparison}

Table~\ref{tab:qualitative_mechanism_comparison} summarizes the performance and implementation characteristics of the considered mechanisms and highlights three distinct regimes. \ac{SLPR} is the most effective, ensuring stable ranging obfuscation (high bias, low variance) with minimal communication and detectability penalties, albeit requiring higher complexity. AM and RD represent attractive low-complexity alternatives. Both achieve deterministic spoofing with low consistency risk, although they incur a moderate communication penalty. Conversely, AN and \ac{ISL} rely on severe signal degradation, causing high variance, poor data-link reliability, and high consistency risk, making them unsuitable for stealthy privacy.

\section{Conclusion}
 This paper presented an initial link-level case study investigating \ac{UE}-side waveform manipulations for location privacy in a single-antenna 5G \ac{NR} uplink scenario, characterizing the trade-offs among localization obfuscation, communication reliability, and detectability. Under the evaluated configuration, our analysis reveals that unstructured perturbations achieve privacy primarily through severe signal corruption, whereas coherent manipulations induce deterministic timing biases with significantly lower communication penalties. Among the considered techniques, the optimized SLPR approach provided the most favorable overall trade-off for this setup, with AM and RD spoofing offering attractive low-complexity alternatives. Crucially, we demonstrated that coherent manipulations can effectively bypass standard network consistency checks, ensuring robust privacy without detection. These findings highlight the potential of waveform-level obfuscation as a promising approach for user privacy preservation in future \ac{ISAC} networks. As future work, extending this evaluation to multi-anchor localization, multi-antenna \ac{gNB} architectures, and diverse scheduling configurations remains a fundamental research direction for privacy-preserving \ac{ISAC} networks.
\begin{table}
\centering
\caption{Qualitative comparison of the considered UE-side privacy mechanisms.}
\label{tab:qualitative_mechanism_comparison}
\footnotesize 
\setlength{\tabcolsep}{4pt} 
\renewcommand{\arraystretch}{0.9} 
\begin{tabular}{@{}lccccc@{}}
\hline
\textbf{Metric} & \textbf{SLPR} & \textbf{ISL} & \textbf{AN} & \textbf{AM} & \textbf{RD}\\
\hline
Bias              & High   & Low    & Low  & High & High \\
Variance          & Low    & High   & High & Low  & Medium  \\
Peak switch       & High   & Low    & Low  & Medium & High \\
BLER              & Low    & High   & High & Medium & Medium \\
Consist. risk     & Low    & High   & High & Low  & Low  \\
Impl. effort      & High   & High   & Low  & Low  & Low  \\
\hline
\end{tabular}
\end{table}

\section*{Acknowledgment}
The authors would like to thank Mahmut K. Ercan for making available the pilot-distortion implementation associated with \cite{ercan2025pilot}. Parts of this implementation were adapted for the optimized pilot-distortion module used in this work.

\balance 
\bibliographystyle{IEEEtran}
\bibliography{IEEEabrv,MyBib}

\end{document}

%% file: acronyms.tex
\newacro{3GPP}[3GPP]{3rd Generation Partnership Project}
\newacro{4G}[4G]{4th generation}
\newacro{5G}[5G]{5th generation}
\newacro{5GS}[5GS]{5G system}
\newacro{6G}[6G]{6th generation}
\newacro{5GAA}[5GAA]{5G automotive association}
\newacro{ADE}[ADE]{average displacement error}
\newacro{ADRF}[ADRF]{Analytics Data Repository Function}
\newacro{AF}[AF]{application function}
\newacro{AI}[AI]{artificial intelligence}
\newacro{AL}[AL]{Alert Limit}
\newacro{AM}[AM]{artificial multipath}
\newacro{AN}[AN]{artificial noise}
\newacro{ANLF}[ANLF]{Analytics Logical Function}
\newacro{AOA}[AOA]{angle-of-arrival}
\newacro{AEF}[AEF]{API exposing function}
\newacro{Apps}[Apps]{applications}
\newacro{A-AOA}[A-AOA]{azimuth angle-of-arrival}
\newacro{AMF}[AMF]{access and mobility function}
\newacro{AOD}[AOD]{angle-of-departure}
\newacro{AoI}[AoI]{area of interest}
\newacro{AP}[AP]{access point}
\newacro{API}[API]{application programming interface}
\newacro{AWGN}[AWGN]{additive white Gaussian noise}
\newacro{B5G}[B5G]{beyond 5G}
\newacro{BBU}[BBU]{Baseband Unit}
\newacro{BLER}[BLER]{block error rate}
\newacro{BER}[BER]{bit error rate}
\newacro{BS}[BS]{base station}
\newacro{C-ITS}[C-ITS]{cooperative intelligent transport systems}
\newacro{CAM}[CAM]{cooperative awareness message}
\newacro{CAPEX}[CAPEX]{capital expenditures}
\newacro{CAPIF}[CAPIF]{Common API Framework}
\newacro{CCF}[CCF]{CAPIF core function}
\newacro{CCO}[CCO]{capacity and coverage optimization}
\newacro{CCRB}[CCRB]{constrained CRB}
\newacro{CCDF}[CCDF]{complementary cumulative distribution function}
\newacro{CID}[CID]{cell identity}
\newacro{CIR}[CIR]{channel impulse response}
\newacro{CIS}[CIS]{continuous intelligent surface}
\newacro{CFAR}[CFAR]{constant false alarm rate}
\newacro{CoO}[CoO]{cell of origin}
\newacro{CPU}[CPU]{central processing unit}
\newacro{CRB}[CRB]{Cram\'er-Rao bound}
\newacro{CNN}[CNN]{convolutional neural network}
\newacro{CP}[CP]{cyclic prefix}
\newacro{CPM}[CPM]{collective perception message}
\newacro{CSI}[CSI]{channel state information}
\newacro{CSI-RS}[CSI-RS]{channel state information reference signal}
\newacro{CSI-RSRP}[CSI-RSRP]{channel state information RSRP}
\newacro{CSI-RSRQ}[CSI-RSRQ]{channel state information RSRQ}
\newacro{CSV}[CSV]{comma-separated value}
\newacro{CU}[CU]{Central Unit}
\newacro{CX}[CX]{customer experience}
\newacro{C-V2X}[C-V2X]{cellular V2X}
\newacro{DAG}[DAG]{direct acyclic graph}
\newacro{DBSCAN}[DBSCAN]{density-based spatial clustering of applications with noise}
\newacro{DCCF}[DCCF]{Data Coordination and Collection Function}
\newacro{DEN}[DEN]{decentralized environmental notification}
\newacro{DFL}[DFL]{device-free localization}
\newacro{DISC}[DISC]{discriminant analysis classification}
\newacro{DL}[DL]{deep learning}
\newacro{DL-AOD}[DL-AOD]{downlink angle-of-departure}
\newacro{DL-PRS}[DL-PRS]{downlink positioning reference signal}
\newacro{DL-TDOA}[DL-TDOA]{downlink time-difference-of-arrival}
\newacro{DMRS}[DMRS]{Demodulation Reference Signal}
\newacro{DNN}[DNN]{deep neural network}
\newacro{DOA}[DOA]{direction-of-arrival}
\newacro{DoS}[DoS]{Denial of Service}
\newacro{DS}[DS]{Delay Spread}
\newacro{DT}[DT]{drive tests}
\newacro{DU}[DU]{Distributed Unit}
\newacro{ECDF}[ECDF]{empirical cumulative density function}
\newacro{ECCDF}[ECCDF]{empirical complementary cumulative distribution function}
\newacro{eCID}[eCID]{Enhanced CID}
\newacro{ECOC}[ECOC]{error/correcting output codes classification}
\newacro{eLCS}[eLCS]{enhanced LCS}
\newacro{EFIM}[EFIM]{equivalent Fisher information matrix}
\newacro{EGNOS}[EGNOS]{european geostationary navigation overlay service}
\newacro{EM}[EM]{electro-magnetic}
\newacro{EMF}[EMF]{electro-magnetic field}
\newacro{eNB}[eNB]{eNodeB}
\newacro{ESMLC}[ESMLC]{enhanced mobile location server}
\newacro{ESPRIT}[ESPRIT]{estimation of signal parameters via rotational invariant techniques}
\newacro{ETSI}[ETSI]{European Telecommunication Standards Institute}
\newacro{FCC}[FCC]{Federal Communication Commission}
\newacro{FDE}[FDE]{final displacement error}
\newacro{FE}[FE]{feature engineering}
\newacro{FIM}[FIM]{Fisher information matrix}
\newacro{FMCW}[FMCW]{frequency modulated continuous wave}
\newacro{FTM}[FTM]{fine time measurement}
\newacro{FLC}[FLC]{fuzzy logic controller}
\newacro{gNB}[gNB]{gNodeB}
\newacro{GAN}[GAN]{generative adversarial networks }
\newacro{GBTs}[GBTs]{gradient boosted trees}
\newacro{GMLC}[GMLC]{gateway mobile location center}
\newacro{GNSS}[GNSS]{global navigation satellite system}
\newacro{GoB}[GoB]{Grid of Beams}
\newacro{GSM}[GSM]{global system for mobile communication}
\newacro{GPS}[GPS]{Global positioning system}
\newacro{GRU}[GRU]{gated recurrent unit}
\newacro{HD}[HD]{high definition}
\newacro{HDBSCAN}[HDBSCAN]{hierarchical DBSCAN}
\newacro{HTTP}[HTTP]{hypertext transfer protocol}
\newacro{ICRB}[ICRB]{intrinsic CRB}
\newacro{ICT}[ICT]{information and communication technology}
\newacro{IIoT}[IIoT]{industrial internet of things}
\newacro{IMU}[IMU]{inertial measurement unit}
\newacro{INS}[INS]{inertial navigation system}
\newacro{IOO}[IOO]{indoor open office}
\newacro{IoT}[IoT]{Internet of Things}
\newacro{IPv6}[IPv6]{internet protocol version 6}
\newacro{IS}[IS]{intelligent surface}
\newacro{ISAC}[ISAC]{integrated sensing and communication}
\newacro{ISL}[ISL]{integrated sidelobe level}
\newacro{ISG}[ISG]{ Industry Specification Group}
\newacro{ISI}[ISI]{inter-symbol interference}
\newacro{ITS}[ITS]{intelligent transport systems}
\newacro{ITS-S}[ITS-S]{ITS station}
\newacro{JRC}[JRC]{Joint Radar and Communication}
\newacro{JSON}[JSON]{JavaScript Object Notation}
\newacro{KPI}[KPI]{key performance indicator}
\newacro{LCS}[LCS]{localization service}
\newacro{LDM}[LDM]{local dynamic map}
\newacro{LDPC}[LDPC]{low density parity check}
\newacro{LM}[LM]{localized measurement}
\newacro{LMF}[LMF]{location management function}
\newacro{LM}[LM]{localized measurement}
\newacro{LOS}[LOS]{line-of-sight}
\newacro{LPP}[LPP]{LTE positioning protocol}
\newacro{LPI}[LPI]{location privacy indication}
\newacro{LSTM}[LSTM]{long short-term memory }
\newacro{LTE}[LTE]{long term evolution}
\newacro{LRF}[LRF]{location retrieval function}
\newacro{LS}[LS]{least squares}
\newacro{MANO}[MANO]{management and orchestration}
\newacro{MBFSN}[MBFSN]{multicast-broadcast single-frequency network}
\newacro{MBS}[MBS]{Metropolitan beacon system}
\newacro{MCC}[MCC]{mobile Country code}
\newacro{MCH}[MCH]{Multicast channel}
\newacro{MCRB}[MCRB]{misspecified CRB}
\newacro{MAE}[MAE]{mean absolute error}
\newacro{MDE}[MDE]{mean distance error}
\newacro{MDT}[MDT]{minimization of drive tests}
\newacro{MFAF}[MFAF]{Messaging Framework Adaptor Function}
\newacro{MITM}[MITM]{Man-in-the-Middle}
\newacro{ML}[ML]{machine learning}
\newacro{MLE}[MLE]{maximum likelihood estimation}
\newacro{MLP}[MLP]{multi-layer perception}
\newacro{mmWave}[mmWave]{millimeter wave}
\newacro{MNO}[MNO]{Mobile Network Operator}
\newacro{MTLF}[MTLF]{Model Training Logical Function}
\newacro{MPC}[MPC]{multi-path component}
\newacro{MIMO}[MIMO]{multiple-input-multiple-output}
\newacro{M-MIMO}[M-MIMO]{massive multiple-input-multiple-output}
\newacro{mm-Wave}[mm-Wave]{millimeter wave}
\newacro{MO-LR}[MO-LR]{mobile originated location request}
\newacro{MPCs}[MPCs]{multiple signal components}
\newacro{MR-DC}[MR-DC]{multi radio - dual connectivity}
\newacro{MSE}[MSE]{mean squared error}
\newacro{MT-LR}[MT-LR]{mobile terminated location request}
\newacro{MU}[MU]{multi-user}
\newacro{MUSIC}[MUSIC]{MUltiple SIgnal Classification}
\newacro{MUT}[MUT]{mean user throughput}
\newacro{multi-RTT}[multi-RTT]{multiple round trip time}
\newacro{NF}[NF]{network function}
\newacro{NCA}[NCA]{neighbourhood component analysis}
\newacro{NEF}[NEF]{network exposure function}
\newacro{NFV}[NFV]{Network Function Virtualization}
\newacro{NG-RAN}[NG-RAN]{next generation-radio access network}
\newacro{NI-LR}[NI-LR]{network induced location request}
\newacro{NMSE}[NMSE]{normalized mean-squared error}
\newacro{NLOS}[NLOS]{non-line-of-sight}
\newacro{NR}[NR]{new radio}
\newacro{NR eCID}[NR eCID]{new radio Enhanced CID}
\newacro{NRPPa}[NRPPa]{new radio positioning protocol a}
\newacro{NSI}{network synthetic image}
\newacro{MCS}[MCS]{modulation and coding scheme}
\newacro{NWDAF}[NWDAF]{Network Data Analytic Function}
\newacro{OFDM}[OFDM]{orthogonal frequency division multiplexing}
\newacro{OPEX}[OPEX]{operational expenditures}
\newacro{OSI}[OSI]{Open System Interconnection}
\newacro{OSS}[OSS]{Operations Support System}
\newacro{OTDOA}[OTDOA]{Observed time-difference-of-arrival}
\newacro{OTFS}[OTFS]{orthogonal time frequency space}
\newacro{PAPR}[PAPR]{peak-to-average power ratio}
\newacro{PCA}[PCA]{principal component analysis}
\newacro{PEB}[PEB]{position error bound}
\newacro{PFL}[PFL]{positioning frequency layers}
\newacro{PNF}[PNF]{Physical Network Function}
\newacro{POI}[POI]{point of interest}
\newacro{PCRB}[PCRB]{posterior CRB}
\newacro{PoTi}[PoTi]{position and time}
\newacro{PL}[PL]{Protection Level}
\newacro{PLMN}[PLMN]{Public Land Mobile Network}
\newacro{PPCA}[PPCA]{Probabilistic PCA}
\newacro{PRB}[PRB]{physical resource block}
\newacro{PRS-RSRP}[PRS-RSRP]{PRS reference signal received power}
\newacro{PRS}[PRS]{positioning reference signal}
\newacro{PSL}[PSL]{positioning service level}
\newacro{PHY}[PHY]{physical layer}
\newacro{PTS}[PTS]{power traffic sharing}
\newacro{PUSCH}[PUSCH]{Physical Uplink Shared Channel}
\newacro{QAM}[QAM]{quadrature amplitude modulation}
\newacro{QoS}[QoS]{quality of service}
\newacro{QPSK}[QPSK]{quadrature phase shift keying}
\newacro{RADAR}[RADAR]{radio detection and ranging}
\newacro{RAIM}[RAIM]{receiver autonomous integrity monitoring}
\newacro{RAN}[RAN]{radio access network}
\newacro{RAT}[RAT]{radio access technology}
\newacro{RE}[RE]{resource element}
\newacro{RedCap}[RedCap]{reduced capability}
\newacro{REM}[REM]{radio environment map}
\newacro{REST}[REST]{REpresentational State Transfer}
\newacro{RCS}[RCS]{radar cross section}
\newacro{RF}[RF]{radio frequency}
\newacro{RIC}[RIC]{RAN Intelligent Controller}
\newacro{RIS}[RIS]{reconfigurable intelligent surface}
\newacro{RMSE}[RMSE]{root mean square error}
\newacro{RNN}[RNN]{recurrent neural network}
\newacro{RRC}[RRC]{radio resource control}
\newacro{RRH}[RRH]{Remote Radio Head}
\newacro{RRM}[RRM]{radio resource management}
\newacro{RS}[RS]{reference signal}
\newacro{RSRP}[RSRP]{reference signal received power}
\newacro{RSRQ}[RSRQ]{reference signal received quality}
\newacro{RSS}[RSS]{received signal strength}
\newacro{RSSI}[RSSI]{received signal strength indicator}
\newacro{RSTD}[RSTD]{received signal time difference}
\newacro{RTK}[RTK]{real time kinematics}
\newacro{RTT}[RTT]{round trip time}
\newacro{RU}[RU]{Radio Unit}
\newacro{SaaS}[SaaS]{software as a service}
\newacro{SAE}[SAE]{society of automotive engineers}
\newacro{SBA}[SBA]{service based architecture}
\newacro{SCI}[SCI]{soft context information}
\newacro{SCS}[SCS]{subcarrier spacing}
\newacro{SDK}[SDK]{software development kit}
\newacro{SEAL}[SEAL]{
Service Enabler Architecture Layer for Verticals}
\newacro{SER}[SER]{Symbol Error Rate}
\newacro{SFI}[SFI]{soft feature information}
\newacro{SI}[SI]{soft information}
\newacro{SINR}[SINR]{signal-to-interference-plus-noise ratio}
\newacro{SIR}[SIR]{signal-to-interference ratio}
\newacro{SL}[SL]{Sidelink}
\newacro{SLA}[SLA]{service-level agreement}
\newacro{SLAM}[SLAM]{simultaneous localization and mapping}
\newacro{SLPP}[SLPP]{Sidelink Positioning Protocol}
\newacro{SLPR}[SLPR]{sidelobe-to-peak level ratio}
\newacro{SNMP}[SNMP]{Simple Network Management Protocol}
\newacro{SNR}[SNR]{signal-to-noise ratio}
\newacro{SMO}[SMO]{Service Management and Orchestration}
\newacro{SINR}[SINR]{signal to interference and noise ratio}
\newacro{SAPTS}[SAPTS]{social-aware PTS controller}
\newacro{SOM}[SOM]{Service Management and Orchestration}
\newacro{SoTA}[SoTA]{state-of-the-art}
\newacro{SPEB}[SPEB]{squared position error bound}
\newacro{SQL}[SQL]{structured query language}
\newacro{SR}[SR]{sensor radar}
\newacro{SRN}[SRN]{sensor radar network}
\newacro{SRS}[SRS]{Sounding Reference Signal}
\newacro{SRS-RSRP}[SRS-RSRP]{SRS reference signal received power}
\newacro{SSB}[SSB]{single-sideband}
\newacro{SSR}[SSR]{state space representation}
\newacro{SS-RSRP}[SS-RSRP]{synchronization-signal-based RSRP}
\newacro{SS-RSRQ}[SS-RSRQ]{synchronization-signal-based RSRQ}
\newacro{SVE}[SVE]{single-value estimation}
\newacro{SVM}[SVM]{support vector machine}
\newacro{TA}[TA]{timing advance}
\newacro{TBS}[TBS]{terrestrial beacon system}
\newacro{TDL}[TDL]{Tap Delay Line}
\newacro{TDOA}[TDOA]{time-difference-of-arrival}
\newacro{TIR}[TIR]{Target Integrity Risk}
\newacro{ToA}[ToA]{time-of-arrival}
\newacro{TOF}[TOF]{time-of-flight}
\newacro{TRP}[TRP]{transmit/receive point}
\newacro{TTA}[TTA]{Time-to-Alert}
\newacro{TTFF}[TTFF]{time to first fix}
\newacro{TTI}[TTI]{transmission time interval}
\newacro{UAV}[UAV]{unmanned aerial vehicle}
\newacro{UDM}[UDM]{unified data manager}
\newacro{UE}[UE]{user equipment}
\newacro{UL-AOA}[UL-AOA]{uplink angle-of-arrival}
\newacro{UL-RToA}[UL-RToA]{uplink relative time-of-arrival}
\newacro{UL-SRS}[UL-SRS]{uplink sounding reference signal}
\newacro{UL-TDOA}[UL-TDOA]{uplink time-difference-of-arrival}
\newacro{UMa}[UMa]{urban macro}
\newacro{UMi}[UMi]{urban micro-cellular}
\newacro{URLLC}[URLLC]{ultra-reliable low-latency communication}
\newacro{UTDOA}[UTDOA]{Uplink time-difference-of-arrival}
\newacro{UWB}[UWB]{ultra-wideband}
\newacro{WIFI}[WIFI]{wireless fidelity}
\newacro{WLAN}[WLAN]{wireless LAN}

\newacro{V2I}[V2I]{vehicle-to-infrastructure}
\newacro{V2P}[V2P]{vehicle-to-pedestrian}
\newacro{V2V}[V2V]{vehicle-to-vehicle}
\newacro{V2X}[V2X]{vehicle-to-everything}
\newacro{VIM}[VIM]{Virtualized Infrastructure Manager}
\newacro{VNF}[VNF]{Virtualized Network Function}
\newacro{VNFM}[VNFM]{VNF Manager}
\newacro{VRU}[VRU]{vulnerable road user}
\newacro{XML}[XML]{(e)Xtensible Markup Language}
\newacro{XR}[XR]{extended reality}
\newacro{Z-AOA}[Z-AOA]{zenith angle-of-arrival}
\newacro{ZSM}[ZSM]{ Zero touch network and Service Management}

%% file: Main.bbl
\begin{thebibliography}{10}
\providecommand{\url}[1]{#1}
\csname url@samestyle\endcsname
\providecommand{\newblock}{\relax}
\providecommand{\bibinfo}[2]{#2}
\providecommand{\BIBentrySTDinterwordspacing}{\spaceskip=0pt\relax}
\providecommand{\BIBentryALTinterwordstretchfactor}{4}
\providecommand{\BIBentryALTinterwordspacing}{\spaceskip=\fontdimen2\font plus
\BIBentryALTinterwordstretchfactor\fontdimen3\font minus \fontdimen4\font\relax}
\providecommand{\BIBforeignlanguage}[2]{{%
\expandafter\ifx\csname l@#1\endcsname\relax
\typeout{** WARNING: IEEEtran.bst: No hyphenation pattern has been}%
\typeout{** loaded for the language `#1'. Using the pattern for}%
\typeout{** the default language instead.}%
\else
\language=\csname l@#1\endcsname
\fi
#2}}
\providecommand{\BIBdecl}{\relax}
\BIBdecl
\renewcommand{\BIBentryALTinterwordstretchfactor}{10}

\bibitem{3gpp.38.211}
3GPP, ``{NR; Physical channels and modulation},'' {3rd Generation Partnership Project (3GPP)}, {Technical Specification (TS)} 38.211, 09 2023, 18.0.0.

\bibitem{3gpp.38.305}
3GPP, ``{NG-RAN; Stage 2 functional specification of User Equipment (UE) positioning in NG-RAN},'' {3rd Generation Partnership Project (3GPP)}, {Technical Specification (TS)} 38.305, 03 2025, 19.1.0.

\bibitem{ayyalasomayajula2023users}
R.~Ayyalasomayajula \emph{et~al.}, ``Users are closer than they appear: Protecting user location from {WiFi} {APs},'' in \emph{Proceedings of the 24th International Workshop on Mobile Computing Systems and Applications}, 2023, pp. 124--130.

\bibitem{checa2020location}
J.~J. Checa \emph{et~al.}, ``Location-privacy-preserving technique for {5G} {mmWave} devices,'' \emph{IEEE Communications Letters}, vol.~24, no.~12, pp. 2692--2695, 2020.

\bibitem{zhang2025privacy}
Y.~Zhang \emph{et~al.}, ``Privacy preservation in mimo-ofdm localization systems: A beamforming approach,'' \emph{IEEE Wireless Communications Letters}, vol.~14, no.~7, pp. 1979--1983, 2025.

\bibitem{tomasin2022beamforming}
S.~Tomasin, ``Beamforming and artificial noise for cross-layer location privacy of e-health cellular devices,'' in \emph{2022 IEEE International Conference on Communications Workshops (ICC Workshops)}.\hskip 1em plus 0.5em minus 0.4em\relax IEEE, 2022, pp. 568--573.

\bibitem{li2024channel}
J.~Li \emph{et~al.}, ``Channel state information-free location-privacy enhancement: Fake path injection,'' \emph{IEEE Transactions on Signal Processing}, vol.~72, pp. 3745--3760, 2024.

\bibitem{li2025delay}
J.~Li \emph{et~al.}, ``Delay-angle information spoofing for channel state information-free location-privacy enhancement,'' \emph{arXiv preprint arXiv:2504.14780}, 2025.

\bibitem{italiano2025holotrace}
L.~Italiano \emph{et~al.}, ``Holotrace: a location privacy preservation solution for mmwave mimo-ofdm systems,'' \emph{arXiv preprint arXiv:2509.23444}, 2025.

\bibitem{gao2024surgical}
K.~Gao \emph{et~al.}, ``Surgical strike on {5G} positioning: Selective-{PRS}-spoofing attacks and its defence,'' \emph{IEEE Journal on Selected Areas in Communications}, vol.~42, no.~10, pp. 2922--2937, 2024.

\bibitem{BarFocPalZanMelBleBia}
S.~Bartoletti \emph{et~al.}, ``Practical blind full-frame replay attacks on {OFDM}-based {ISAC} systems,'' \emph{IEEE Journal on Selected Areas in Communications}, vol.~44, pp. 4239--4253, 2026.

\bibitem{FocZanPalBiaBar:26}
G.~Focarelli \emph{et~al.}, ``Positioning security in {5G} and beyond: Model and detection of physical layer threats,'' \emph{IEEE Transactions on Wireless Communications}, vol.~25, pp. 1048--1061, 2026.

\bibitem{ercan2025pilot}
M.~K. Ercan \emph{et~al.}, ``Pilot distortion design for {ToA} obfuscation in uplink {OFDM} communication,'' \emph{arXiv preprint arXiv:2510.24223}, 2025.

\bibitem{zhang2024privacy}
Y.~Zhang \emph{et~al.}, ``Privacy preservation in delay-based localization systems: Artificial noise or artificial multipath?'' in \emph{GLOBECOM 2024 - 2024 IEEE Global Communications Conference}, 2024, pp. 2755--2760.

\end{thebibliography}
